\documentclass[preprint,5p,times,twocolumn]{elsarticle}
\usepackage{verbatim}
\usepackage{listings}
\usepackage{multirow}
\usepackage{graphicx} 
\usepackage{lineno}
\usepackage{array}
\usepackage{hyperref} 
\usepackage{graphicx}
\usepackage{amsmath,amssymb,amsthm}
\usepackage{rotating}
\usepackage{subfigure} %

\newcommand{\ca}{capacitance }

\def\mnu{$\langle m_{\nu} \rangle$~}

\def\BBz{$\beta\beta(0\nu)$~}
\def\BBzn{$\beta\beta(0\nu)$}

\def\BBd{$\beta\beta(2\nu)$~}
\def\BBdn{$\beta\beta(2\nu)$}
\def\BB{$\beta\beta$~}
\def\BBn{$\beta\beta$}

\def\ca{$\sim$}

\def\dot{$\cdot$}

\def\teod{TeO$_2$~}

\def\be{\begin{equation}}
\def\ee{\end{equation}}

\def\per{$\times$}
\def\Dlt{$\Delta$}

\begin{document}

\begin{frontmatter}

\title{Experimental searches of neutrinoless double beta decay}

\author{Oliviero Cremonesi}

\address{INFN - Sez. Milano Bicocca, Milano I-20126 - Italy}

\begin{abstract}
	Neutrinoless double decay  (\BBzn) is a unique probe for  lepton number conservation and neutrino properties. It allows to investigate the Dirac/Majorana nature of the neutrinos and their absolute mass scale (hierarchy problem) with unprecedented sensitivity. A number of experiments are presently under preparation to cover the quasi-degenerate region of the neutrino mass spectrum. Improved sensitivities are however required to sound the so-called inverted hierarchy region. This is a real challenge faced by a number of new proposed projects, based either on large expansions of the present experiments or on new ideas top improve the technical performance or reduce the background contributions. A review of the most relevant ongoing experiments is given. The most relevant parameters contributing to the experimental sensitivity are discussed and a critical comparison of the future projects is proposed. 
\end{abstract}


\end{frontmatter}


\section{Neutrinoless Double Beta Decay}
\label{NDBD}
	Neutrino oscillation experiments have shown that neutrinos are massive particles that mix through the PMNS matrix to give rise to the flavor eigenstates. 
	Moreover, recent results from reactor experiments have shown that all the three mixing angles are different from zero, opening a new window on the search for CP violations on the leptonic sector. This is a very strong demonstration that the Standard Model of electroweak interactions is incomplete and that new Physics beyond it must exist. 
	However, two very important neutrino properties are still missing: their nature and the absolute scale of their masses. Neutrinoless double beta decay seems presently the most effective way to find an answer to these questions. 
	In fact, present techniques for direct measurements of the electron antineutrino mass (the only to guarantee a model-independent approach) can only probe the quasi-degenerate region ($\delta$m $\ll$ m), while the much more sensitive consmological inferences suffer from a heavy model dependance. All these experimental approaches provide however complementary pieces of information and a common effort is compulsory.
	
Double Beta Decay (\BBn) is a rare nuclear process in which a parent nucleus (A,Z) decays to a member (A,Z+2) of the same isobaric multiplet with the simultaneous emission of two electrons. 
	Among the possible \BB modes two are of particular interest, the 2$\nu$ mode (\BBdn) $^A_ZX \to ^A_{Z+2}X + 2e^- + 2\overline{\nu}$, which observes the lepton number conservation and it is allowed by the Standard Model (SM) of electro-weak interactions, and the 0$\nu$ mode (\BBzn) $^A_ZX \to ^A_{Z+2}X + 2e^-$ which violates the lepton number by two units and occurs only if neutrinos are their own antiparticles.
\BBz is one of the most powerful tools to test neutrino properties. Indeed it can exist only if neutrinos are Majorana particles. Furthermore, thanks to its strict relation with neutrino masses it can provide important constraints on the neutrino mass scale. 

When mediated by the exchange of a light virtual Majorana neutrino, the \BBz rate can be expressed as
	\begin{equation}\label{eq:rate}
	[ T_{1/2}^{0\nu}]^{-1} = 
	G^{0\nu}|M^{0\nu}|^2{\vert\langle m_{\nu} \rangle\vert^2}/m_e^2
	\end{equation}
where $G^{0\nu}$ is the (exactly calculable) phase space integral, $|M^{0\nu}|^2$ is the nuclear matrix element and \mnu is a (coherent) linear combination of the neutrino masses
	\begin{equation}\label{eq:mnu}\begin{split}
	\langle m_{\nu} \rangle \equiv \sum_{k=1}^{3}\vert	U_{ek}^L \vert ^2m_ke^{i\phi _k } \\
	\simeq c_{12}^2 c_{13}^2 m_1+s_{12}^2 c_{13}^2 e^{i\alpha_1} m_2+s_{13}^2 e^{i\alpha_2}m_3 
	\end{split}\end{equation}
The last equality holds for small neutrino masses and $\alpha_1$ and $\alpha_2$ are the so-called neutrino Majorana phases whose presence in (\ref{eq:mnu}) implies that cancellations are possible. 
On the other hand \BBz represents a unique possibility to measure the neutrino Majorana phases.

	Altogether, the observation of \BBz and the accurate determination of the \mnu would  establish definitely that neutrinos are Majorana particles, fixing their mass scale and  providing a crucial contribution to the determination of the absolute neutrino mass scale.
	However, even in the case that forthcoming \BBz experiments would not observe any decay, important constraints can be obtained.  Indeed, assuming that neutrinos are Majorana particles, a negative result in the 20-30 meV range for \mnu would definitely rule out the inverse ordering thus fixing the neutrino hierarchy problem. 
	On the other hand, if future oscillation experiments would demonstrate the inverted ordering of the neutrino masses, a failure in observing \BBz at a sensitivity of 20-30 meV would show that neutrinos are Dirac particles.

	As can be easily deduced from eq.~(\ref{eq:rate}), the derivation of the only neutrino relevant parameter \mnu from the experimental \BBz results  requires a precise knowledge of the transition Nuclear Matrix Elements M$^{0\nu}$(NME) for which many (unfortunately often conflicting) evaluations are available in the literature. 
	Significative improvements have been obtained recently which have reduced the spread in the computed NME values within a factor 2-3\cite{faess11}, even if SM calculations are still systematically smaller than the others. Such an agreement does not guarantee by itself the correctness of the calculations but the convergence of the results from very different methods can hardly be a chance. However new ideas on how to approach NME calculations and extensive and independent cross-checks of the calculation results seem compulsory.
	Useless to say that from a purely experimental point of view, the spread in the available NME calculations causes a lot of confusion in the comparison of results and sensitivities of the different experiments. 
	While waiting for a general agreement, the only possibility is to refer to a single (arbitrarily chosen) calculation, or to a ``Physics Motivated Average'' (PMA) set of NME values~\cite{gome10} or disentangling the uncertainty intervals according to the different calculations\cite{crem10}.
	In order to preserve correlations and allow a (relative) comparison between the sensitivities of \BBz experiments,  we will refer here to a single calculation \cite{ibm2} which has the advantage of being available for all the nuclei of interest. 
	\section{The experiments}
	Most of the experiments on  \BBz are based on counter methods for the direct observation of the two electrons emitted in the decay. They aim at collecting the limited available information (sum of the electron energies, single electron energy and angular distributions, identification and/or counting of the daughter nucleus) and are usually classified in {\em inhomogeneous} (when the observed electrons originate in an external sample) and {\em homogeneous} experiments (when the source  of \BB's serves also as detector).
	Both approaches are characterized by attractive features even if homogeneous experiments have provided so far the best results and characterize most of the future proposed projects.
	In most cases the different \BB modes are separated simply on the basis of the different distribution expected for the electron sum energies: a continuous bell distribution for \BBdn, and a sharp line at the transition energy for \BBz. Therefore, a good energy resolution is a very attractive experimental feature. 

	Experimental evidence for several \BBd decays has been provided using the measured two-electron sum energy spectra, the single electron energy distributions and the event topology\cite{barab11,exo12,kzen12}. 
	On the other hand, impressive progress has been obtained during the last years also in improving \BBz half-life limits for a number of isotopes (Tab. \ref{tab:BBpres}). The best results are still maintained by experiments based on the use of isotopically enriched HPGe diodes (Heidelberg-Moscow\cite{hmosc01} and IGEX\cite{aalse02}) but two other experiments reached comparable sensitivities: NEMO3\cite{klang09,barab05} at LSM and CUORICINO at LNGS\cite{arnab08}. 
	\begin{table}
	\caption{\label{tab:BBpres} Best reported results on \BBd  and \BBz processes and most relevant \BB parameters. Limits are at 90\% CL. \mnu are computed using NME and phase space factors from \cite{ibm2} and \cite{ips12} respectively.}
	\centering
	\begin{tabular}{@{}lcccc@{}}
	\hline
	Isotope 	& T$_{1/2}^{2\nu}$\cite{barab11}           & T$_{1/2}^{0\nu}$  & \mnu &Q  \\ 
	 	& (10$^{19}$y)  & (10$^{24}$y) & (eV) &(MeV)   \\ 
	\hline
	$^{48}$Ca          & $(4.4^{+0.6}_{-0.5})$	& $>0.0014 $\cite{ogawa04} & 14 &4.27\\
	$^{76}$Ge          & $(150\pm 10)$ & $>19$\cite{hmosc01}& 0.44& 2.04\\
	                   & & $22.3^{+4.4}_{-3.1}$\cite{klapd08}& 0.4\\
	                   & & $>15.7$\cite{aalse02}& 0.5\\
	$^{82}$Se          & $(9.2 \pm 0.7)$ & $>0.36$ \cite{klang09}& 1.9 &2.995\\
	$^{96}$Zr  		   & $(2.3\pm0.2)$ & $>0.0092$\cite{klang09}& 15 & 3.35\\
	$^{100}$Mo         & $(0.71\pm0.04)$& $>1.1$\cite{klang09}& 1.0 & 3.034\\
	$^{116}$Cd         & $(2.8\pm 0.2)$& $>0.17$\cite{danev03}& 3.5 &2.802\\
	$^{130}$Te         & $(68^{+12}_{-11})$& $>2.8$\cite{cuor10}& 0.6 &2.527\\
	$^{136}$Xe         & $>81$\cite{gavri00} & $>0.45$\cite{exo12} & 0.3 &2.479\\
	$^{150}$Nd         & $(13.3^{+4.5}_{-2.6})$ & $>0.0018$\cite{barab05} & 21 &3.367\\
	$^{238}$U          & $(220\pm50)$ & $>0.0036$\cite{barab05}\\ 
	\hline
	\end{tabular} 
	\end{table}
	Evidence for a \BBz signal has also been claimed\cite{kkdh01} (and later maintained~\cite{klapd08}) by a small subset of the HDM collaboration at LNGS with $T^{0\nu}_{1/2}=2.23^{+0.44}_{-0.31}\times 10^{25}$ y. The result is based on a sophisticated re-analysis of the HDM data heavily relying on pulse shape analysis and artificial neural network algorithms aiming at identifying the \BBz signal whlile reducing the background contributions. 
	Such a claim has raised a lot of criticism but cannot be dismissed out of hand. On the other hand, none of the existing experiments can rule out it, and the only certain way to confirm or refute it is with additional sensitive experiments. In particular, next generation experiments should easily achieve this goal.

	The performance of a \BBz experiment is usually expressed in terms of a detector {\em factor of merit} (or sensitivity), defined as the process half-life corresponding to the maximum signal n$_B$ that could be hidden by the background fluctuations at a given statistical C.L. 
	At 1$\sigma$ level (n$_B$=$\sqrt{BTM\Delta}$), one obtains:
	\begin{equation}\begin{split}
	\label{eq:sensitivity}
	F_{0\nu} = \tau^{Back.Fluct.}_{1/2}= \ln 2~N_{\beta\beta}\epsilon\frac{T}{n_B} = \\
	= \ln 2\times \frac{x ~\eta ~ \epsilon ~ N_A}{A} \sqrt{ \frac{ M ~ T }{B ~ \Delta} } ~ (68\% CL)
	\end{split}\end{equation}  
	where B is the background level per unit mass and energy, M is the detector mass, T is the measure time, $\Delta$ is the FWHM energy resolution, N$_{\beta\beta}$ is the number of \BB decaying nuclei under observation, $\eta$ their isotopic abundance, N$_A$ the Avogadro number, A the compound molecular mass, $x$ the number of \BB atoms per molecule, and $\epsilon$ the detection efficiency.
Actually B never scales exactly with the detector mass but this approximation is usually reasonable and has a physical justification.

The case when the background level B is so low that the expected number of background events in the region of interest along the experiment life is of order of unity (B \dot M\dot T \dot \Dlt \ca O(1)) deserves particular attention. In these case one generally speaks of ''zero background'' (0B) experiments, a condition met by a number of upcoming projects. In these conditions, eq. (\ref{eq:sensitivity}) can no more be used and a good approximation to the sensitivity is given by
	\begin{eqnarray}
	\label{eq:0sensitivity}
	F_{0\nu}^{0B} = 
	\ln 2~N_{\beta\beta}\epsilon\frac{T}{n_L} \nonumber 
	= \ln 2\times \frac{x ~\eta ~ \epsilon ~ N_A}{A} 
	\frac{ M ~ T }{n_L}
	\end{eqnarray}  

	where $n_L$ is a constant depending on the chosen CL and on the actual number of observed events. 
  The most relevant feature of equation (\ref{eq:0sensitivity}) is that F$_{0\nu}^{0B}$ does not depend on the background level or the energy resolution and scales linearly with the sensitive mass M and the measure time T. 
Since T is usually limited to a few years and $\Delta$ is usually fixed (meaning that for a given experimental technique it is usually difficult to get sizable improvements), the 0B condition translates to B\dot M \ca O(1)/\Dlt \dot T). This means that for a given mass M there always exists a threshold for B below which no further improvement of the sensitivity is obtained or, alternatively, that it can be useless to reduce at will the background level without a corresponding increase of the experimental mass. A well designed experiment has therefore to match the condition $B\cdot M \gtrsim 1/\Delta\cdot T$.
For most of the next generation high resolution calorimeters this corresponds to $B_T \simeq \frac {1}{10\cdot M}$ or $B_T \simeq 10^{-4}$ for a O(1t) experiment.

Despite its simplicity, equations (\ref{eq:sensitivity}) and  (\ref{eq:0sensitivity}) have the unique advantage of emphasizing the role of the essential experimental parameters: mass, measuring time, isotopic abundance, background level and detection efficiency. 
Actually most of the criteria to be considered when optimizing  the design of a new \BBz experiment follow  directly from the above equations: i) a well performing detector (e.g. good energy resolution and time stability) giving the maximum number of informations (e.g. electron energies and event topology); ii) a reliable and easy to operate detector technology requiring a minimum level of maintenance (long underground running times); iii) a very large (possibly isotopically enriched) mass, of the order of one ton or larger; iv) an effective background suppression strategy. These criteria are actually being pursued by all the next generation experiments. Unfortunately, they are often conflicting and their simultaneous optimisation is rarely possible. 
On the other hand, they don't take into account important details like the shape of the expected signal or of the background and can't be used to analyze the case of very low statistics. In these cases a more sophisticated Monte Carlo approach is needed.
A series of new proposals has been boosted in recent years by the renewed interest in \BBz following neutrino oscillation results. The ultimate goal is to reach sensitivities such to allow an investigation of the inverted hierarchy (IH) of neutrino masses (\mnu\ca10-50 meV). From an experimental point of view this corresponds however to active masses of the order of 1~ton (or larger) with background levels as low as \ca~1 c/keV/ton/y. A challenge that can hardly be faced by the current technology. Phased programs have been therefore proposed in USA and Europe\cite{matrix,aspera}.

Next generation experiments are all characterized by hundred kg detectors and 1-10 c/kev/ton background rates. Their goal is to select the best technology and approach the IH region. A restricted list of some of the most advanced forthcoming \BBz projects is given in Table \ref{tab:BBres}. 
Very different classification schemes can of course be adopted for them. They are usually based on the different strategies adopted to improve the \BBz sensitivity: experimental approach, mass, energy resolution, background discrimination technique, granularity and track reconstruction, etc. 
\begin{center}\begin{table*}
\caption{\label{tab:BBres} List of some of the most developed \BBz projects. 5 years sensitivity at 90\% C.L. Experimental phases are indicated as running (R), progress (P) or development (D). \mnu values are calculated using NME and phase space factors from \cite{ibm2} and \cite{ips12} respectively. Asterisk signals 0B condition. B' is the background per unit of isotope mass in units of 10$^{-3}$~counts/keV/kg/y. }
	\centering
	\begin{tabular}{@{}llclcccccc@{}}
	\hline
			&Isotope 	&Mass	&B' &FWHM &Lab	&Status&Start &S$_{5}^{0\nu}$&\mnu\\
			& 			&[kg]	& &[keV] &[keV]		&		&	  &[10$^{26}$y]&[meV] \\ 
	\hline
	CUORE\cite{cuore}	&$^{130}$Te	&206 &29 &5	&LNGS	&P	&2014 &2.1 &73\\
	GERDA I\cite{gerda}	&$^{76}$Ge	&18	&23 &4.5 &LNGS	&R	&2012	&1.1 &184\\
	GERDA II		&			&40	&1.2 &3 &		&D	&		&2.1* &133\\
	MJD\cite{majorana}&			&30	&1.2 &3 &SUSEL	&P	&2014	&2.6* & 67\\
	EXO\cite{exo12}		&$^{136}$Xe	&200 &1.9 &100	&WIPP	&R	&2011 & 1.2 & 115\\
	SuperNEMO\cite{supernemo}&$^{82}$Se	&100-200&0.08&120 &LSM	&D	&2013-2015 & 0.8 & 90\\
	KamLAND-Zen\cite{kzen12}&$^{116}$Cd&400 &9	&248 &Kamioka&R	&2011	& 0.33 &220\\
			&			&1000	&9 & 248 &		&D	&2013-2015	& 0.59 & 164\\
	SNO+\cite{sno}	&$^{150}$Nd	&44	&1.8	&229 &Sudbury&D	&2013	& 0.08 &310\\
	SNO+ II	&	&131	&1.8	&303 &Sudbury&D	&2013	& 0.08 &310\\
	NEXT\cite{next}	&			&100 &0.2	&13 & Canfranc		&D	&2014	& 5* & 56\\ 
\hline
\end{tabular}\end{table*}\end{center}
In general, three broad classes can generally be identified: i) arrays of calorimeters with excellent energy resolution and improved background suppression methods (e.g. GERDA, MAJORANA) or based on unconventional techniques (e.g. CUORE); ii) detectors with generally poor energy resolution but topology reconstruction (e.g. EXO, SuperNEMO); iii) experiments based on suitable modifications of an existing setup aiming at a different search (e.g. SNO+, KAMLAND). 
In some cases technical feasibility tests are required, but the crucial issue is still the capability of each project to pursue the expected background suppression. Different estimates of the expected B levels are usually based on the extrapolation of real measurements to the final experimental conditions or on the Monte Carlo simulations based on more or less realistic expectations. The former are usually more reliable especially when based on the results of medium size detectors (protorypes). The expected sensitivities are listed in Tab.\ref{tab:BBres}. Here the measured values (``Measured'') are distinguished from realistic projections (``Reference'') and most optimistic expectations (``Improved''). Experiments entering the 0B regime are also indicated. 
Although all proposed projects show interesting features for a next generation experiment, only few of them are characterized by a reasonable technical feasibility within the next few years. 

MAJORANA and GERDA belong to the class of the high energy resolutions calorimeters and are both phased programs representing large scale extensions of past successful experiments on $^{76}$Ge \BBzn. Background control is based upon a careful choice of the setup materials and of very effective radiation shields. Active reduction based on new detector design for single site event identification represent the new frontier and are presently gathering most of the experimental efforts. In both cases this is accomplished by means of p-type ``Broad Energy'' isotopically enriched germanium diodes (or ``BEGe'').

	CUORE\cite{cuore} is a very large extension of the \teod bolometric array concept pioneered by the Milano group at the Gran Sasso Laboratory since the eighties. CUORE consists of a rather compact cylindrical structure of 988 cubic natural \teod crystals of 5~cm side (750~g) operated at a temperature of ~10~mK. The expected energy resolution is \ca5~keV FWHM at the \BBz transition energy (\ca 2.53~MeV). The expected background level is of the order of \ca 0.01 c/keV/kg/y. The expected 5y sensitivity is $2.1\times10^{26}$ y allowing a close look at the IH region of neutrino masses. CUORE is presently under construction at LNGS. 

	Thanks to the bolometer's versatility, alternative options with respect to \teod are also possible. In particular, promising results have been recently obtained with scintillating bolometers which are particularly effective in identifying the dangerous alpha background from the surface of the detector setup\cite{scib10}. 

	Gas and liquid TPC's represent another aspect of the homogeneous approach in which the limited resolution is the most relevant limitation while scalability and geometrical reconstruction are the most evident advantages.
	EXO ({\it Enriched Xenon Observatory}) is a challenging project based on a large mass (\ca~1--10 tons) of isotopically enriched (85\% in  $^{136}$Xe) Xenon. A sizable prototype experiment with a Xe mass of 200 kg (80\% $^{136}$Xe), has been deployed at WIPP since summer 2009 and has recently published  a lower limit of 1.6 10$^{25}$ yr on the $^{136}$Xe  \BBz half-life\cite{exo12}. Further improvements on energy resolution and background are still expected while the experiment is approved to run for 4 more years. 
	Expected to operate at LSC, NEXT is a mainly Spanish project based on the use of a high pressure Xe gas TPC for a better energy resolution and topological signature for a powerful background rejection\cite{next}. It aims at a phased program starting with a 100 kg. Smaller scale prototypes have been already built and operated successfully providing excellent results on energy resolution\cite{next12}.

	New developments have been  proposed more recently in order to exploit two successful experiments on neutrino oscillation like SNO and KamLAND. 
	SNO+ is pursuing the goal of studying $^{150}$Nd with 0.78 to 2.24 tons of natural Neodimium dispersed in a balloon filled with a liquid scintillator. 
	
	The same concept is applied by KAMLAND-Zen, in which a large masses of $^{136}$Xe is dispersed in the liquid scintillator. Proposed in 2009, the program has started in September 2011 with 320 kg of 90\% enriched $^{136}$Xe. Also this experiment has recently presented the first results characterized by an unexpected large background level in the ROI. A strong effort to identify its origin and reduce its effects is presently ongoing.  

	The proposed Super-NEMO experiment is the only project based on an inhomogeneous approach. It is an extension of the successful NEMO3 concept, properly scaled  in order to accommodate \ca100 kg of $^{82}$Se foils spread among 20 detector modules. The expected energy resolution is 7\% FWHM (12\% in NEMO-3) to improve the signal detection efficiency from 8\% to 40\% and reduce the \BBd contribution. The projected background is \ca $3.5\times 10^{-4}$ c/keV/kg. A demonstrator (single module) is presently fully funded to be operated in the current NEMO3 site. 
	\begin{figure*}
	\caption{\label{fig:BBcomp} Comparison between a selected number of running, proposed or under construction \BBz projects. Iso-sensitivity lines are plotted in brown while green arrows show the direction of fastest increase of the sensitivity. 
}
	\centering\includegraphics[width=0.85\textwidth]{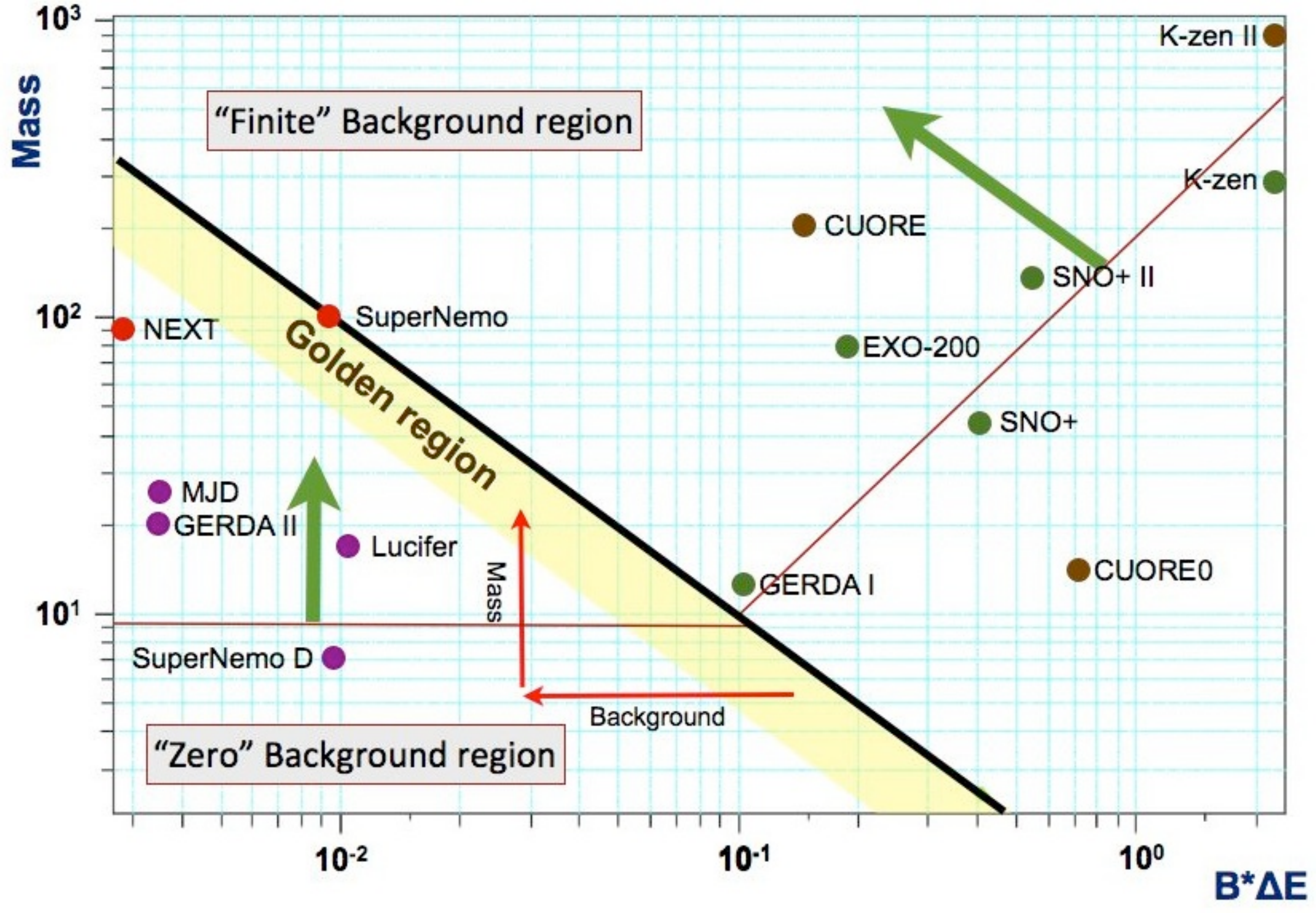}%
	\end{figure*}
\section{Critical comparison}
	A nice feature of equations (\ref{eq:sensitivity}) and  (\ref{eq:0sensitivity}) is that the relevant experimental parameters appear in them always in the combinations M\per{T} and B\per${\Delta}$. 
	This suggests a nice way to compare the performance of the proposed experiments as points on a plane (M\per{T}, B\per${\Delta}$). 
	Indeed, when using a log-log scale (Fig.~\ref{fig:BBcomp}), the condition for the transition to the 0B regime (M\per{T})\per(B\per${\Delta}$)\ca~O(1) is simply a line with slope -1  (transition line). On the other hand the iso-sensitivity curves are also lines: horizontal in the case of 0B regime (where the increase is linear with mass) and with slope +1 in the normal background region (where the increase is with the square root of (M\per{T}) over (B\per${\Delta}$)). 
	It should also be stressed that in the 0B region no increase of sensitivity is associated, for a given mass, to a decrease of the background. The goal of the 0B experiments should be therefore, for a given background level, to increase the experimental mass in order to approach the so called ``golden'' region near the transition line between the two background regimes, a condition which is not met by any of the proposed low background demonstrators. 
	Actually they are all very far from the transition line and this means that none of them actually exploits the tremendous effort for reducing the background level and, moreover, that since the expected total number of events in the ROI is much less than unity, they can only provide a model dependent demonstration of the reached background level through an integration over large energy intervals on the basis of proper background models

	Due to the different increase of the sensitivity in the normal and 0B regions, the iso-sensitivity lines do not match at the transition line and a quantitative comparison is effective only separately in each region. However the plot allows a quick and critical comparison between the different proposed experiments, weighted on the sensitivity parameters and outlining the best strategy for future improvements.
	
	\section{Conclusions}
	Neutrino oscillation results have stimulated a renewed interest in the experimental study of neutrino properties. In this framework, neutrinoless \BB decay is a unique tool to verify the Majorana nature of the neutrino and can provide important information on the neutrino mass scale and intrinsic phases, unavailable to the other neutrino experiments.  
	An international effort is supporting a phased \BBz program based on a number of newly proposed experiments aiming at reaching sensitivities to test the inverted neutrino mass hierarchy. Three next generation experiments have already started data taking while other will soon be ready.
	The success of the upcoming \BBz program strongly depends on the true capability of the proposed projects to reach the required background levels in the ROI.


\begin{thebibliography}{99}
	\bibitem{furry39} W.H.~Furry, {\it Phys. Rev.} {\bf 56} (1939) 1184.
	\bibitem{faess11} A.~Faessler, nucl-th/1104.3700v1.
	\bibitem{gome10} J.J.~Gomez-Cadenas et al, {\it Rivista del Nuovo Cimento} {\bf 35} (2012) 29–98; hep-ex/1010.5112v3.
	\bibitem{crem10} O.~Cremonesi,  hep-ex/1002.1437.
	\bibitem{ibm2} J.~Barea and F.~Iachello, {\it Phys. Rev. C} 79 (2009) 044301.
	\bibitem{ips12} J.~Kotila and F.~Iachello, {\it Phys. Rev. C} 85 (2012) 034316.
	\bibitem{barab11} A.S.~Barabash, {\it Phys.Atom.Nucl.} 74 (2011) 603.
	\bibitem{ogawa04} I.~Ogawa et al., {\it Nucl. Phys.} A {\bf 730} (2004) 215
	\bibitem{hmosc01} H.V.~Klapdor-Kleingrothaus et al. (HM Coll.), {\it Eur. Phys. J.} A {\bf 12}, 147 (2001) and hep-ph/0103062
	\bibitem{klapd08} H.V.~Klapdor-Kleingrothaus, {\it Phys. Scr.} T {\bf 127} (2006) 40; H.V.~Klapdor-Kleingrothaus and I.V.~Krivosheina,  {\it Modern Physics Letters} A {\bf 21} (2006) 1547.
	\bibitem{aalse02} C.E.~Aalseth et al., {\it Phys. Rev.} D {\bf 65} (2002) 092007
	\bibitem{klang09} K.~Lang (NEMO3 coll.), Proc. fo the WIN09 conference, Perugia September 2009, Italy.
	\bibitem{danev03} F.A.~Danevich et al., {\it Phys. Rev.} C {\bf 68} (2003) 035501
	\bibitem{cuor10} E. Andreotti et al., {\it Astropart.Phys.} 34 (2011) 822; nucl-ex/1012.3266.
	\bibitem{gavri00} J.M.~Gavriljuk et al., {\it Phys. Rev.} C {\bf 61}, 035501 (2000).
	\bibitem{exo12} J.~Kotila and F.~Iachello, {\it Phys. Rev. C} 85 (2012) 034316.
	\bibitem{barab05} A.S.~Barabash, {\it Phys. At. Nucl.} {\bf 68} (2005) 414.
	\bibitem{kzen12} A.~Gando et al. hep-ex:1201.4664.	%
	\bibitem{arnab08} C.~Arnaboldi et al., {\it Phys. Rev.} C {\bf 78} (2008) 035502
	\bibitem{kkdh01} H.V.~Klapdor-Kleingrothauset al., {\it Mod. Phys. Lett.} A16 (2001) 2409, hep-ph/0201231.
	\bibitem{matrix} S.J.~Freedman, B.Kayser: ''The Neutrino Matrix: DNP / DPF / DAP / DPB Joint study on the future of Neutrino Physics''; physics/0411216.
	\bibitem{aspera} Astroparticle Physics: the European strategy. http://www.aspera-eu.org.
	\bibitem{cuore} C.~Arnaboldi et al., {\it Nucl. Instr. Meth. A} 518 (2004) 775; C.~Arnaboldi et al., hep-ex/0501010; I.C.~Bandac et al., {\it J. Phys. Conf. Ser.} 110 (2008) 082001.
	\bibitem{gerda} I.~Abt	et	al., hep-ex/0404039; J.~Janicsko-Csathy, {\it Nucl.	Phys.	B	(Proc.
	Suppl.)} 188 (2009) 68.
	\bibitem{majorana} C.E.~Aalseth et al., hep-ex/0201021; Majorana White Paper, nucl-ex/0311013; V.E.	Guiseppe, nucl-ex/0811.2446.
	\bibitem{supernemo} A.S.~Barabash, {\it Phys. At. Nucl.} 79 (2004) 10; A.S.~Barabash, {\it Czech. J. Phys.} 52 (2002) 567, nucl-ex/0203001; A.S.~Barabash, {\it Phys. At. Nucl.} 67 (2004) 438; E.~Chauveau, {\it AIP Conf. Proc.} 1180 (2009) 26.
	\bibitem{sno} J.~Maneira, Proc. Neutrino Oscillation Workshop, Conca Specchiulla, Italy, September 4-11, 2010)
	\bibitem{next}F.~Granena et al., hep-ex/0907.4054v1; JJ.~Gomez-Cadenas et al.,physics.ins-det/1210.0341.	
	\bibitem{next12} V.~Alvarez et al., JINST 7, T06001 (2012); physics.ins-det/1202.0721; physics.ins-det/1211.4474.
	\bibitem{scib10} C.~Arnaboldi et al, nucl-ex/1011.5415;  C.~Arnaboldi et al., {\it Astropart.Phys.} 34 (2011) 344, nucl-ex/1006.2721; C.~Arnaboldi, {\it Astropart.Phys.} 34 (2010) 143, nucl-ex/1005.1239.
\end{thebibliography}
\end{document}